\documentclass[12pt]{article} 

\usepackage{times} % uncomment for Times Roman;
\usepackage{wrapfig,lipsum,booktabs,amssymb,amsmath}
\usepackage{setspace}
\usepackage{bm} % If you want bold math, but witt the bloat of AMS packages
\usepackage{color} % doh
\usepackage{hyperref} % LaTeX cross references become hyperlinks in pdf output
\usepackage[font=small, belowskip=-14pt,aboveskip=0pt]{caption}
\usepackage{graphicx} % For figure insertion
\usepackage{vmargin} % To set page size
\usepackage{titlesec}
\usepackage{multirow}
\usepackage[normalem]{ulem}
\usepackage{braket}
\usepackage{cancel}
\usepackage{makecell}
\titlespacing{\section}{5pt}{\parskip}{-\parskip}
\titlespacing{\subsection}{3pt}{\parskip}{-\parskip}
\titlespacing{\subsubsection}{2pt}{\parskip}{-\parskip}
 
\usepackage[style=nature,citestyle=numeric-comp, sorting=none, backend=biber, maxbibnames=9]{biblatex}
\hypersetup{colorlinks=true, citecolor=black,linkcolor=blue,urlcolor=blue}
\DeclareCaptionFont{blue}{\color{blue}}
\usepackage{etoolbox}
\patchcmd{\thebibliography}{\section*{\refname}}{}{}{}
\usepackage{authblk}

\setpapersize{USletter}

\setmarginsrb{1in}{1in}{1in}{1in}{0pt}{0mm}{0pt}{0mm}

\begin{document}

\title{Skyrmion Quantum Diode Prototype: Bridging Micromagnetic Simulations and Quantum Models}

\author[1]{Haowen Yang}
\author[1]{Gerald Bissell}
\author[1]{Han Zhong}
\author[1]{Peter Van Kirk}
\author[1]{Tiger Cao}
\affil[1]{Department of Electrical and Computer Engineering, University of Florida, Gainesville, FL 32611, USA}
\author[2]{Pengcheng Lu}
\affil[2]{Department of Physics, University of Michigan, Ann Arbor, MI 48109, USA}

\author[1]{Yingying Wu\footnote{Corresponding author: yingyingwu@ufl.edu}}
\date{\vspace{-5ex}}
\maketitle

\begin{abstract}
Magnetic skyrmions are topologically protected spin textures known for their robustness against perturbations. Their topological stability makes them robust information carriers, ideal for tackling a key challenge in quantum computing: creating reliable, one-way links between different types of qubits. In this proof-of-concept study, we introduce a novel device—the skyrmion quantum diode—based on skyrmion qubits. Our approach combines classical micromagnetic simulations, achieving skyrmion diameters as small as 3 nm, with quantum circuit models inspired by superconducting qubits.
In this work, we demonstrate:
(i) unidirectional skyrmion transport via the skyrmion Hall effect in asymmetric junctions, spanning length scales from 20 nm down to 3 nm;
(ii) potential compatibility with flux-tunable quantum architectures; and
(iii) preliminary insights into anharmonicity in skyrmion-based qubit systems.
These results establish both the operational feasibility and the scaling behavior necessary for a hybrid skyrmion–quantum platform. Our work outlines a path toward integrating skyrmion-based quantum components into practical device architectures, enabling low-dissipation, unidirectional quantum information transport. This capability is crucial for scalable quantum computing, spintronic logic, and hybrid quantum systems, and opens opportunities for chip-scale, pump-free isolators and directional quantum links that enhance readout fidelity, reduce cryogenic load, and support modular skyrmion–superconducting processors.
\end{abstract}

\section*{Introduction}
A central challenge in superconducting qubits is their susceptibility to decoherence and operational imperfections, which continue to hinder scalability in quantum information processing. Although substantial progress has been made in extending coherence times and improving gate fidelities through advances in materials, device design, and control techniques~\cite{clarke2008superconducting,abughanem2025superconducting}, preserving robust quantum states in large-scale circuits remains a significant obstacle. As circuit complexity increases, the impact of environmental noise, energy dissipation, crosstalk, and fabrication-induced disorder becomes more pronounced, leading to error accumulation and performance degradation. These issues not only limit the reliability of quantum operations but also place stringent demands on error correction overhead, thereby constraining the practicality of scalable superconducting quantum architectures. Recently, significant advances have been made in implementing nonreciprocal elements, such as superconducting diodes~\cite{nadeem2023superconducting,zhong2025twisted}. For example, superconducting qubit platforms have demonstrated quantum transport in a Su–Schrieffer–Heeger (SSH) chain of tunable couplers~\cite{zhao2023engineering}, where directional excitation transfer can be realized using only a small number of lossy elements. Such SSH diodes represent an important milestone toward the development of compact, on-chip components that enforce nonreciprocal signal flow and suppress backscattering in quantum circuits. Twisted superconducting diode was also recently demonstrated~\cite{zhong2025twisted} in NbSe$_2$ few layers with a twisting angle of 1$^\circ$. These superconducting diodes can serve as Josephson junctions in the quantum circuit. The quantum simulation was carried out in a transmon circuit to show the fidelity enhancement on the increased diode efficiency. The theoretically diode efficiency can reach 100\%. However, in current experiments, it was limited to 70\%~\cite{hou2023ubiquitous}. 

To complement these circuit-based approaches, magnetic skyrmions~\cite{nagaosa2013topological, wu2020neel, wu2022van, zhang20242d} offer a promising alternative qubit platform. These particle-like magnetic solitons, stabilized by the Dzyaloshinskii-Moriya interaction (DMI) or magnetic frustration, exhibit exceptional nanoscale stability, inherent robustness against defects, and efficient current-driven mobility. The size of magnetic skyrmions observed in experiments varies from $\sim$ 3 nm to several micrometers. Unlike conventional qubits, which rely on precise parameter control and isolation from the environment, skyrmions benefit from intrinsic topological resilience, providing a natural mechanism for suppressing sensitivity to local perturbations. This built-in robustness offers a potential route to mitigating the high error rates that currently limit superconducting architectures. Recent theoretical and experimental work suggests that skyrmions can host quantum degrees of freedom, such as quantized helicity~\cite{xia2022nonlinear, psaroudaki2021skyrmion, zhong2025integrating, wu2025spin}, enabling their use as quantum two-level systems. For example, information is stored in the quantum degree of helicity in the skyrmion qubit, and the logical states can be adjusted by electric and magnetic fields, offering a rich operation regime with high anharmonicity~\cite{psaroudaki2021skyrmion}. These skyrmion-based qubits combine operational tunability, positioning them as a compelling platform for scalable quantum information processing. Note topological protection is rigorously defined in the continuum limit and can be weakened in discrete spin systems, particularly for skyrmion qubits composed of only a few spins. These skyrmion qubits have so far been explored mainly through theoretical models, calculations, and simulations~\cite{petrovic2025colloquium}, and many aspects of their behavior remain to be investigated and fully understood.

Here, we propose a hybrid approach, using a skyrmion-based quantum diode to provide this missing nonreciprocal function directly on-chip. The skyrmion's topological charge guarantees it behaves as a robust, particle-like information carrier. By guiding it through a
T-shape nanotrack, its motion becomes inherently directional, creating a one-way channel for quantum information. 
Micromagnetic simulations were employed to demonstrate the operation of a skyrmion Hall effect-based diode at 3 nm skyrmion sizes. Building on this, we conducted energy-level structure analysis of skyrmion qubits along with fidelity simulations to evaluate their quantum performance. We further carried out comparative circuit-level modeling between superconducting and skyrmion-based qubit diodes to assess their relative advantages.

\section*{Results}
\subsection*{Micromagnetic Simulation of Skyrmion Diode Scaling}
\label{sec:micromag}
%\subsubsection*{Model, geometry, and parameter sweep}
We simulate current-driven N\'eel skyrmions in a T-shaped asymmetric nanotrack acting as a skyrmion diode~\cite{feng2022skyrmion}. Unless otherwise stated, material parameters are adopted from hard magnet case, e.g. between Fe$_3$GeTe$_2$ values~\cite{hu2024room} and CoFeB thin films case~\cite{xu2021effects}:
\[
M_s=580\times10^3~\mathrm{A/m},\quad
A_\mathrm{ex}=15\times10^{-12}~\mathrm{J/m},\quad
D=3.0~\mathrm{mJ/m^2}.
\]
Perpendicular anisotropy is swept over
\(
K_u\in[0.8,\,1.5]\times10^6~\mathrm{J/m^3},
\)
and the track dimensions are co-scaled to target stable skyrmion core diameters from \(\sim20~\mathrm{nm}\) down to \(\sim3~\mathrm{nm}\).
Damping is set to \(\alpha=0.1\).
We employ MuMax3 with a cell size chosen \(<\!l_\mathrm{ex}/10\) throughout; for the smallest devices we use \(\Delta x=\Delta y=\Delta z\leq0.1~\mathrm{nm}\), ensuring the skyrmion core is resolved across many cells even as its diameter approaches a few nanometers.
The magnet is initialized in the uniform \(+\hat z\) state with a single skyrmion seeded at the input arm.
A spin–orbit torque (SOT) drive is applied as a uniform in-plane current density \(J\) (0.2$\times$10$^{12}$ A/m\(^2\)) to drive the translation of skyrmions.
\begin{figure}[ht]
    \centering
    \includegraphics[width=\textwidth]{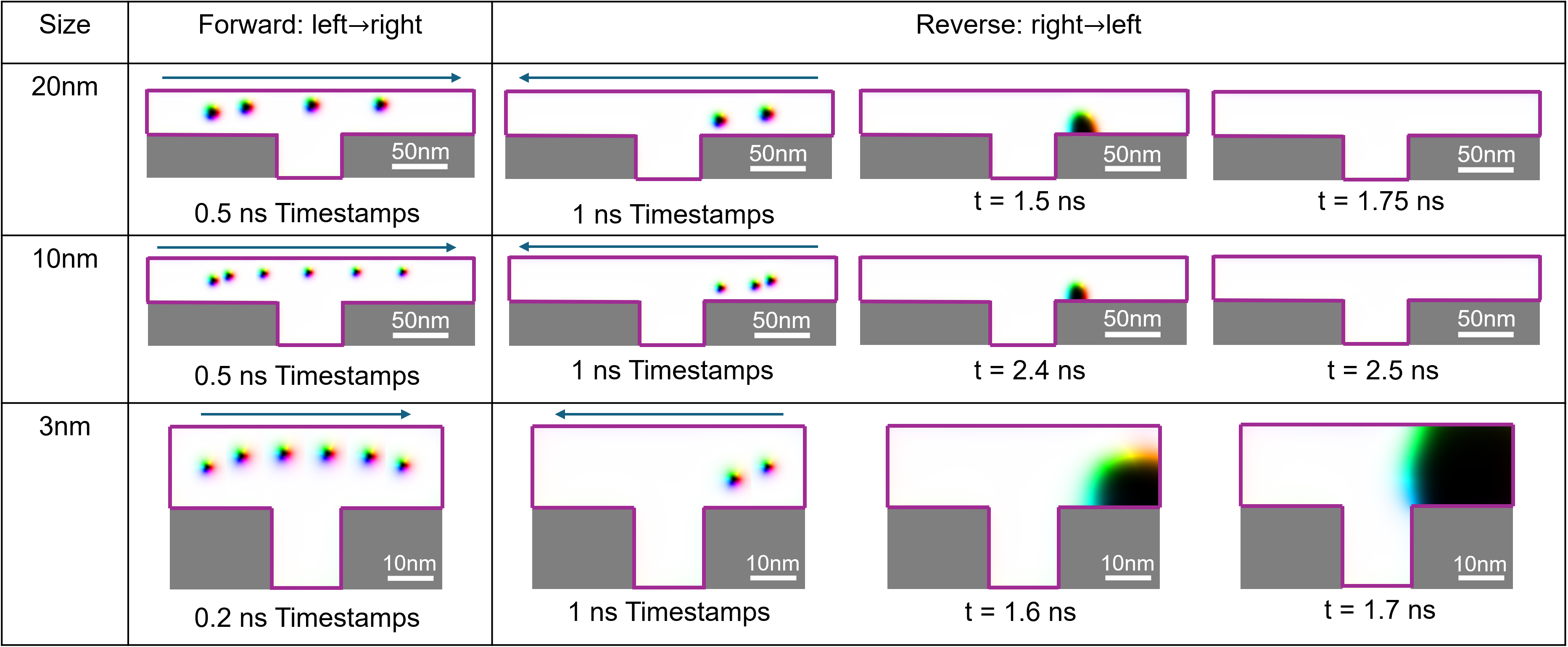}
    \caption{Micromagnetic simulation snapshots showing forward (left~$\rightarrow$~right) and reverse (right~$\rightarrow$~left) propagation of N\'eel skyrmions through an asymmetric T-junction diode for three target core diameters: $\sim$20~nm, $\sim$10~nm, and $\sim$3~nm. In forward bias, the skyrmion Hall effect steers skyrmions into the widening side of the junction, enabling transmission. In reverse bias, Hall deflection toward the narrowed side; in all types of skyrmions, this leads to reflection back into the injection arm. Time Stamps indicate intervals between frames for each size case.}
    \label{fig:skyrmion_sizes}
\end{figure}

%\subsubsection*{Simulation protocol}
To test the diode's nonreciprocal behavior, we performed two simulations for each $(K_u, \text{geometry})$ parameter set, with the outcome determined by visual inspection of the skyrmion's trajectory. In each case, a single skyrmion is initialized and relaxed within the main horizontal track before being propelled by a uniform spin-current density~$J$.

\begin{enumerate}
    \item[(i)] \emph{Forward Bias:} The skyrmion is initially placed in the left section of the track. We observe its successful passage through the junction into the right arm, as illustrated in Fig.~\ref{fig:skyrmion_sizes}.

    \item[(ii)] \emph{Reverse Bias:} The skyrmion is initially placed in the right section. Under the same drive current, we observe that the skyrmion Hall effect causes it to be deflected and reflected by the junction, preventing transmission.
\end{enumerate}

To scale from 20 nm skyrmions to 3 nm, we increased $K_u$ from $0.8$ to $1.5$~MJ/m$^3$ (with fixed $M_s$, $A_\mathrm{ex}$, $D$) and proportionally shrank the track and throat reduces the skyrmion radius monotonically. Across $\sim20$~nm down to $\sim3$~nm cores we observe one-way transport: in forward bias the skyrmion transits the junction in $\tau_{\mathrm{fwd}}\!\sim\!0.8$--$2.5$~ns (for the $\sim3$~nm case, $\tau_{\mathrm{fwd}}\!\approx\!1.2$~ns), while in reverse bias it approaches the asymmetric throat and returns to the injection arm within $\tau_{\mathrm{rev}}\!\lesssim\!3$~ns (no transmission). It is also noted that in our simulation, we noticed the smaller skyrmion, $\sim3$~nm case, is unable to move laterally in $\pm x$ direction. Therefore, we applied a spin polarized current to drive the skyrmion. The skyrmion dynamic, in this case driven by spin-transfer torques (STT) with uniform in-plane current density \(J\) (0.4$\times$10$^{12}$ A/m\(^2\)), also demonstrates a huge hall effect and leading the skyrmion to hit the bottom wall and deflection toward the injection node. Thus, the nonreciprocal decision occurs on sub-nanosecond scales.
%To scale from 20 nm skyrmions to 3 nm, we increased $K_u$ from $0.8$ to $1.5$~MJ/m$^3$ (with fixed $M_s$, $A_\mathrm{ex}$, $D$) and proportionally shrank the track and throat reduces the skyrmion radius monotonically. Across $\sim20$~nm down to $\sim3$~nm cores we observe one-way transport: in forward bias the skyrmion transits the junction in $\tau_{\mathrm{fwd}}\!\sim\!0.8$--$1.3$~ns (for the $\sim3$~nm case, $\tau_{\mathrm{fwd}}\!\approx\!1.2$~ns), while in reverse bias it approaches the asymmetric throat and returns to the injection arm within $\tau_{\mathrm{rev}}\!\lesssim\!0.3$~ns (no transmission). Thus, the nonreciprocal decision occurs on sub-nanosecond scales.

%\paragraph{Numerical note.}
We use in-plane cells as small as $\Delta x=\Delta y=0.1$~nm for the smallest skyrmions. Because the continuum model degrades when feature sizes approach a few nanometers, results at $\sim3$~nm should be interpreted as a limiting trend.

Note that micromagnetic simulations are fundamentally classical and cannot capture quantum coherence or superposition effects. While they accurately describe skyrmion spin textures, collective coordinates, and classical nonreciprocal transport, no quantization of these modes is incorporated in our current model. Consequently, the proposed ``quantum diode” functionality is not directly simulated; the classical model primarily guides device design and mode selection in the nanoscale, while fully quantum behavior would require a separate framework beyond classical micromagnetics.

%\subsubsection*{Numerical resolution and validity}
For the smallest devices the skyrmion diameter approaches $3$~nm. To avoid under-resolving the core (spurious pinning and non-physical dynamics when only a handful of cells span the texture), we use an in-plane cell size as small as $\Delta x=\Delta y=0.1$~nm. This choice ensures $\gtrsim\!20$ cells across the skyrmion diameter and suppresses grid-induced artifacts.

This mesh is stricter than standard micromagnetic guidelines ($\Delta \lesssim \min[\ell_{\rm ex},\Delta_{\rm DW}]/5$); with $\quad A_\mathrm{ex}=15$~pJ/m and $M_s=580$~kA/m we have $\ell_{\rm ex}\!\approx\!8.4$~nm, and with $K_u=0.8$--$1.5\times 10^6$ J/m$^3$ the wall-width parameter is $\Delta_{\rm DW}\!\approx\!3.0$--$4.3$~nm. Thus $\Delta \lesssim 0.6$--$0.8$~nm would already satisfy continuum resolution; we over-resolve to maintain $\ge$10--20 cells across the core at 3~nm.

We emphasize that such fine meshing is a simulation choice: micromagnetics is a continuum theory and does not become physically more accurate below atomic length scales.

%\subsubsection*{``Approach \& return'' at the asymmetric junction: mechanism}

To create a skyrmion, commonly used approaches~\cite{da2025neuromorphic,tchoe2012skyrmion,hrabec2017current,kang2016skyrmion} include spin-polarized current, current pulses, and defect- or geometry-assisted nucleation. In our T-shaped track, we drive the skyrmion left$\to$right or right$\to$left with a spin Hall effect (SHE)-induced damping-like SOT. The dynamics are well captured by a rigid-core Thiele picture,
\begin{equation}
  \mathbf{G}\times\mathbf{v}
+ \alpha_{\mathrm G}\,\underline{\underline{D}}\,\mathbf{v}
- \nabla U(\mathbf{r})
= \mathbf{F}_{\mathrm{DL}}(j),
  \label{eq:thiele_sot}
\end{equation} 
where $\mathbf G\propto Q\hat z$ is the gyrotropic (Magnus) vector fixed by the skyrmion charge $Q$, $\underline{\underline D}$ is the dissipative tensor, $U(\mathbf r)$ is the geometric confinement potential of the junction (Fig. \ref{fig:fidelity_maps}a), and $\mathbf F_{\mathrm{DL}}(j)$ is the SOT drive (the field-like term is negligible here). We discuss further with separate forward and reverse motion. 

%\paragraph{Reverse bias (approach \& return).}
The Magnus term $\mathbf G\times\mathbf v$ bends the trajectory toward one edge. In the reverse direction, this Hall deflection leads into the ``hard'' (thin) side of the T-throat where curvature and demagnetizing/DMI boundary charges steepen $U(\mathbf r)$. If $j<j_c^{\rm rev}$, the longitudinal component of $\mathbf v$ is extinguished by $-\nabla U$, after which the combined Magnus and edge forces rotate $\mathbf v$ and the skyrmion returns to the injection arm rather than transmitting \cite{thiele1973steady}.

%\paragraph{Forward bias (one-way transmission).}
In the forward direction, the same Hall deflection steers into the ``soft'' (wide) side, so $-\nabla U$ assists motion through the junction. We therefore observe one-way transfer for all sizes studied (from $\sim20$~nm down to $\sim3$~nm cores), provided $j$ exceeds the depinning threshold yet remains below the annihilation current at the throat.

These Hall–geometry interactions at constrictions are consistent with prior analyses of skyrmion motion in confined and asymmetric tracks \cite{iwasaki2013current,jiang2017direct,jung2021magnetic}.
The T-shaped, Hall-assisted junction functions as a one-way skyrmion diode from $\sim20$~nm down to $\sim3$~nm with sub-nanosecond transit—i.e., a diode based on nanoscale skyrmions in micromagnetic simulations. 
In the following, its potential applicability as an actual quantum diode is explored by first investigating the Josephson junction (flux-tunable transmon) it would interface with.

\subsection*{Quantum Simulation with Skyrmion Qubits}
\label{sec:quantum-sim}

\begin{figure*}[t]
    \centering
    \includegraphics[width=\textwidth]{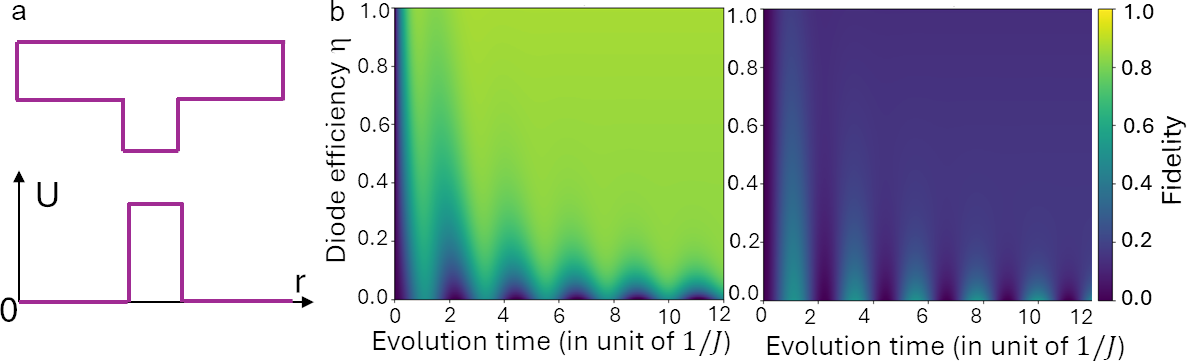}
     \caption{Fidelity mapping. (a) Schematic geometric potential. (b) Fidelity values for a single skyrmion qubit under varying diode efficiency $\eta$.
    Middle: forward fidelity $F_{L\to R}$ (initial $\lvert 0\rangle$, measured in $\lvert 1\rangle$).
    Right: reverse fidelity $F_{R\to L}$ (initial $\lvert 1\rangle$, measured in $\lvert 0\rangle$).
    Horizontal axis: evolution time $t$ (in units of $1/J$). Vertical axis: diode efficiency $\eta$.}
    \label{fig:fidelity_maps}
\end{figure*}

To explore the effect of diode efficiency on qubit transport fidelity, we model a single skyrmion qubit as a two-level open quantum system with Hamiltonian
\begin{equation}
H = J\,\sigma_x + \delta\,\sigma_z ,
\label{eq:Hamil}
\end{equation}
where $J$ is the coherent exchange coupling between qubit states and $\delta$ represents an effective detuning term arising from skyrmion asymmetry or local anisotropy\cite{pan2024magnon}.

Diode effect, or directional transport, is not inherent in this Hamiltonian. Following the physical motivation in the provided supplemental material (see in supplementary), a non-reciprocal interaction can be achieved through engineered dissipation. We model this phenomenologically by introducing a directional loss channel via an amplitude-damping collapse operator,
\begin{equation}
C = \sqrt{\gamma(\eta)}\,\lvert 0\rangle\langle 1\rvert,
\qquad
\gamma(\eta) = \eta\,\gamma_{\text{max}},
\end{equation}
so that increasing $\eta$ corresponds to stronger one-way relaxation from the excited to ground state, representing higher diode efficiency.

Forward and reverse fidelity was computed via time evolution of the master equation:
\begin{align}
F_{L\to R}(t,\eta) &= \big|\langle 1 | \psi(t;\eta,\lvert 0\rangle)\rangle\big|^2,\\
F_{R\to L}(t,\eta) &= \big|\langle 0 | \psi(t;\eta,\lvert 1\rangle)\rangle\big|^2,
\end{align}
where $\psi(t;\eta,\psi_0)$ denotes the qubit state evolved under the dissipative Liouvillian dynamics for diode efficiency $\eta$.
%\subsection{Metric and Procedure}

As $\eta$ increases as shown in Fig. \ref{fig:fidelity_maps}b, the damping term $\gamma(\eta)$ suppresses coherent oscillations and drives the system asymmetrically: forward fidelity (left panel) decreases with stronger diode effect, while reverse fidelity (right panel) increases since the decay process favors $\lvert 0\rangle$ population. The contrast between the two maps thus captures the diode’s nonreciprocal influence on qubit-state evolution.

To clarify, the physical interpretation of these intermediate fidelities reflects quantum-state mixing within the internal helicity degree of freedom rather than partial transport in real space. In the classical diode simulations, the skyrmion behaves as a rigid particle whose transport is binary, either fully transmitted or fully blocked, so no intermediate outcomes exist. In contrast, the quantum model describes the evolution of the helicity qubit, not the classical particle position, and the fidelity reflects the overlap between the evolving quantum state and a target helicity eigenstate. Thus, fidelities between 0 and 1 do not represent partial or probabilistic transmission of a skyrmion through the junction. Instead, they correspond to coherent superpositions of the helicity basis states induced by the same chiral asymmetry responsible for diode behavior in the classical limit. The ``intermediate fidelity” therefore reflects quantum-state mixing within the internal helicity degree of freedom rather than partial transport in real space.

%\subsubsection*{Relation to T-shaped Skyrmion Diode Geometry}

In the micromagnetic system, the parameter $\eta$ represents the diode efficiency of a T-shaped skyrmion track that permits motion in one direction while partially reflecting or dissipating motion in the reverse direction. This geometric asymmetry—originating from variations in interfacial DMI, exchange stiffness, or notch geometry—translates into an effective relaxation rate for the quantum model. Hence, higher $\eta$ values emulate stronger nonreciprocity, corresponding to a skyrmion path that is nearly one-way. This analogy allows us to treat skyrmion transport as a dissipative process acting on a qubit’s excited-state amplitude.

Although this single-qubit model does not simulate full spatial transport, it effectively captures the impact of diode-induced asymmetry on qubit dynamics and coherence.

\subsection*{Enhancing Skyrmion Qubit Anharmonicity}\label{sec:qubit}

%\subsubsection*{Helicity-based skyrmion qubits}

Magnetic skyrmions are nanoscale, topologically nontrivial spin textures stabilized in chiral and frustrated magnets by a competition between exchange, anisotropy, and DMI; their emergent electrodynamics and robustness have made them central to spintronics and information processing proposals \cite{nagaosa2013topological}. In addition to their center-of-mass and radius, skyrmions possess an internal helicity $\phi_0$---a $2\pi$-periodic angle describing the in-plane rotation of spins about the core---which distinguishes Bloch-type from N\'eel-type textures and can be tuned by material symmetry, dipolar interactions, and fields \cite{kong2024diverse, tatarskiy2024direct}. When skyrmions are reduced to nanometric scales and operated at low temperatures, quantum dynamics of their collective coordinates becomes relevant; path-integral treatments and effective-coordinate quantization predict discrete spectra for these modes, providing a route to qubit encodings based on collective variables rather than microscopic spins \cite{psaroudaki2017quantum}.

A helicity skyrmion qubit encodes $\lvert 0\rangle,\lvert 1\rangle$ in the lowest quantum states of $\phi_0$ confined by a periodic pinning potential $V(\phi_0)$ that is well captured by even harmonics such as $K_2\cos 2\phi_0$ (favoring Bloch or N\'eel orientations) and, when present, weak odd harmonics from external bias. Theoretical proposals show that helicity can serve as a controllable quantum degree of freedom, with gate operations implemented by time-dependent electric fields, spin currents, or microwave magnetic fields; this establishes skyrmion helicity as a viable platform for quantum logic \cite{psaroudaki2021skyrmion}. In frustrated magnets, in particular, the helicity exhibits near-degeneracy that supports qubit architectures and universal quantum computation schemes at nanoscale, highlighting materials routes to strong nonlinearity and controllability without superconducting circuitry.

From a modeling standpoint, the helicity behaves as a quantum rotor. $\phi_0$ is a continuous $2\pi$-periodic coordinate with conjugate angular momentum $S_z\in\mathbb{Z}$. The effective Hamiltonian takes the generic form $H=\bar\kappa_z S_z^2 + V(\phi_0)$, where $\bar\kappa_z$ sets the moment of inertia and $V(\phi_0)$ derives from crystalline anisotropy, DMI symmetry, and magnetostatic energy after projecting onto the collective mode. Quantization of this rotor leads to discrete intrawell levels, a tunnel-split ground doublet in a double-well potential, and anharmonic spacings between $\omega_{01}$ and $\omega_{12}$---the key resource for addressable qubit operations \cite{psaroudaki2017quantum}. Although the analogy is not one-to-one, this use of a cosine-like potential to obtain addressable transitions echoes the role of the Josephson cosine in superconducting transmon, where the intrinsic anharmonicity of the potential enables selective control \cite{koch2007charge}.

Recent experiments and materials studies further motivate helicity-based encodings by demonstrating control over helicity in multilayers and the breadth of skyrmionic textures stabilized by DMI and dipolar interactions across platform classes \cite{hassan2024dipolar,hoffmann2017antiskyrmions}. This helicity-based encoding discussed in this section is intended for skyrmions realized in frustrated or centrosymmetric magnets, rather than in conventional chiral magnets governed by interfacial or bulk DMI.

Here we clarify the connection between the real-space skyrmion diode and the helicity-based qubit model. While the diode operation treats the skyrmion as a rigid quasiparticle governed by translational dynamics, the qubit description focuses on the quantized internal helicity mode. These degrees of freedom become coupled in realistic devices because the same chiral asymmetry, arising from DMI gradients and interfacial engineering, that induces nonreciprocal transport also breaks the rotational symmetry of the skyrmion texture, generating a helicity-dependent effective potential. As the skyrmion moves across this asymmetric landscape, its internal configuration experiences position-dependent distortions that map the real-space asymmetry onto the helicity energy spectrum, yielding the double-well potential used for the qubit. Thus, the classical diode effect and the quantum helicity states represent two manifestations of the same underlying chiral engineering, and we have revised the manuscript to provide a more explicit and intuitive explanation of this unified physical mechanism.

%\subsubsection*{Introducing diode efficiency}

 To begin with, the conventional helicity--skyrmion qubit Hamiltonian \cite{psaroudaki2021skyrmion} is written in the collective-coordinate (quantum rotor) form for the helicity angle $\phi_0$ with conjugate $S_z\in\mathbb{Z}$:
\begin{align}
H_0 &= \bar{\kappa}_z S_z^2-\bar{h}_z S_z+V_0(\phi_0), \label{eq:H0}\\
V_0(\phi_0) &= K_2\cos\!\big(2\phi_0\big)-e_z\cos\!\phi_0. \label{eq:V0}
\end{align}
Here $\bar{\kappa}_z S_z^2$ sets the rotor kinetic scale (moment of inertia), $K_2\cos(2\phi_0)$ is the leading even-harmonic pinning that distinguishes Bloch- from N\'eel type and creates a double well with period $\pi$, and the weak odd harmonic $-e_z\cos\phi_0$ represents symmetry-breaking biases such as small in-plane fields. Even harmonics determine the barrier and intrawell curvature, while odd harmonics weakly select a well without setting the nonlinearity.

Attaching a skyrmion racetrack diode to the qubit adds a steady, direction-selective skyrmion flow that modifies the time-averaged magnetic environment in the pad through stray fields, exchange-mediated boundary conditions, and spin--orbit torques. After coarse-graining over the racetrack dynamics, the correction that couples most strongly to helicity must respect the $\phi_0\to\phi_0+\pi$ symmetry associated with flipping all in-plane spins. This symmetry excludes a net odd-harmonic contribution and singles out an even harmonic as the leading term. Therefore the diode contributes an additional $\cos(2\phi_0)$ component whose amplitude scales with the efficiency of one-way transport.

\begin{figure}[h]
\centering
\includegraphics[width=1.0\linewidth]{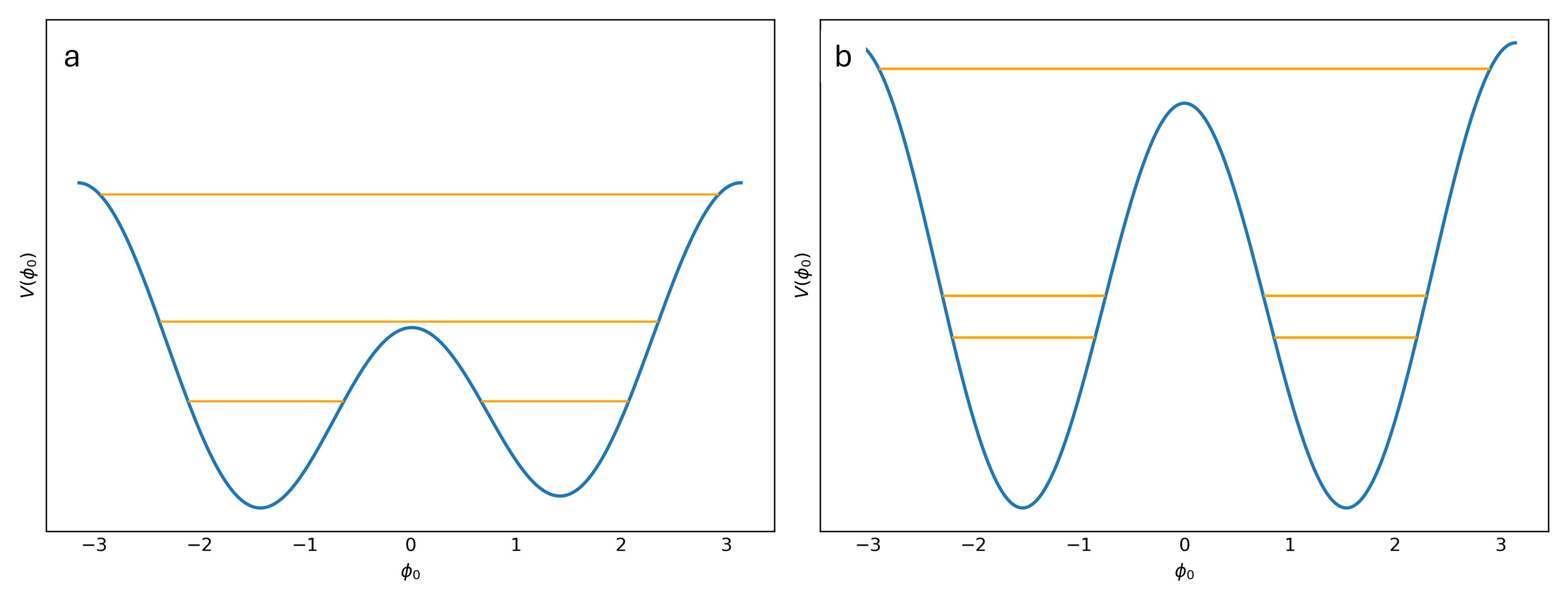}
\caption{\label{fig:energylevels} Comparison of the helicity–skyrmion qubit potential and quantized levels with and without a diode-efficiency term. \textbf{(a)} Conventional double-well potential with the lowest eigen-energies drawn as constrained horizontal segments (only where \(V(\phi_0)\le E_n\)). \textbf{(b)} Diode-modified potential \(V(\phi_0)=K_2^{\mathrm{eff}}\cos(2\phi_0)-\bar{E}_z\cos\phi_0\) with \(K_2^{\mathrm{eff}}=\eta K_2\), where the efficiency factor \(\eta\) renormalizes the \(\cos(2\phi_0)\) term and yields greater level-spacing nonuniformity---i.e., stronger anharmonicity---relative to (a).}
\end{figure}

We parameterize this by introducing a dimensionless diode efficiency $\eta\in[0,1]$ and writing the effective even-harmonic coefficient as
\begin{equation}
K_2^{\mathrm{eff}}(\eta)=K_2+\eta\,K_2^{(\mathrm{D})}, \label{eq:K2eff_general}
\end{equation}
where $K_2^{(\mathrm{D})}\ge 0$ is a device- and geometry-dependent scale that quantifies how strongly the rectified skyrmion flux enhances the helicity pinning. This linear-response form captures the experimentally observed proportionality between rectified skyrmion flow and static modifications of nearby magnetic energies in the operating regime. In our simulations we use the simplest normalization in which the diode-controlled piece sets the scale,
\begin{equation}
K_2^{\mathrm{eff}}(\eta)=\eta\,K_2, \label{eq:K2eff_eta}
\end{equation}
which is equivalent to the more general form after absorbing any baseline pinning into $K_2$.
Substituting $K_2\to K_2^{\mathrm{eff}}(\eta)$ into $H_0$ yields the diode-augmented Hamiltonian
\begin{equation}
H(\eta)=\bar{\kappa}_z S_z^2-\bar{h}_z S_z+\eta\,K_2\cos\!\big(2\phi_0\big)-e_z\cos\!\phi_0. \label{eq:H_eta}
\end{equation}
No further structural changes are introduced: the diode term is Hermitian, preserves the $\pi$-periodicity of the helicity sector, and couples in the $m$-basis exactly as the conventional $\cos(2\phi_0)$ term (matrix elements between $\lvert m\rangle$ and $\lvert m\pm2\rangle$ simply acquire the replacement $K_2\mapsto \eta K_2$. In the angle picture this corresponds to a barrier modulation that deepens the wells and steepens intrawell curvature as $\eta$ increases while leaving the minima approximately fixed. The result in Fig. \ref{fig:energylevels} is a controlled enhancement of spectral anharmonicity, with $\lvert\omega_{12}-\omega_{01}\rvert$ increasing under a single, materials-native parameter $\eta$ that is set at the circuit level by racetrack current and geometry.

%\subsection{Discussion}

Enhancing the anharmonicity $\Delta\omega=\omega_{12}-\omega_{01}$ of the helicity–skyrmion qubit directly improves gate selectivity, readout contrast, and robustness to noise. A larger $|\Delta\omega|$ increases the detuning between the computational transition and higher levels, which suppresses population leakage during fast pulses and enables shorter gates or lower drive power at fixed error rates—an established principle in weakly nonlinear superconducting qubits that carry over to rotor-based qubits by the same spectral-isolation logic \cite{krantz2019quantum, hyyppa2024reducing, motzoi2009simple}. In dispersive readout, greater intrinsic nonlinearity improves level distinguishability by enlarging state-dependent shifts and reducing spectral crowding, thereby raising measurement signal-to-noise and lowering measurement-induced errors \cite{blais2021circuit}. Deeper intrawell confinement associated with an increased even-harmonic barrier also narrows the ground-state wavefunction and reduces susceptibility to low-frequency parameter drift, conceptually akin to operating at bias points that minimize first-order dephasing channels \cite{koch2007charge}. Altogether, engineering larger $|\Delta\omega|$—in our case by scaling the even-harmonic barrier via the diode efficiency $\eta$—yields a more isolated two-level manifold with reduced leakage, cleaner readout, and improved dephasing performance without additional cryogenic control hardware \cite{krantz2019quantum}.

\subsection*{Skyrmion-Transmon Tuning}

Having established through micromagnetic and quantum-circuit simulations that the skyrmion diode exhibits robust nonreciprocity and preserves coherence in forward transport, we now outline how such a diode can be integrated with a superconducting qubit. In this section we present a concrete hybrid architecture in which a flux-tunable transmon is inductively coupled to the diode’s output arm, derive the relevant flux-dependent resonance relations, and discuss implications for isolation and scalability.

The transmon qubit is a capacitively-shunted variant of the Cooper-pair box (CPB) engineered to suppress charge-noise sensitivity while retaining sufficient anharmonicity for qubit operations \cite{koch2007charge}. It consists of a single Josephson junction (or, alternatively, a SQUID loop for tunability) with Josephson energy $E_J$ shunted by a large capacitor that sets the charging energy $E_\mathrm{C} = \frac{e^2}{2C_\Sigma}$ (with $C_\Sigma$ representing the total island capacitance). By operating with an increased $\frac{E_J}{E_C}$ ratio, the device exponentially suppresses charge dispersion, reducing dephasing from offset-charge fluctuations by several orders of magnitude compared to the CPB \cite{koch2007charge, schreier2008suppressing}.

%\subsection{Skyrmion-transmon coupling}

\begin{figure}[h]
\centering
\includegraphics[width=.5\linewidth]{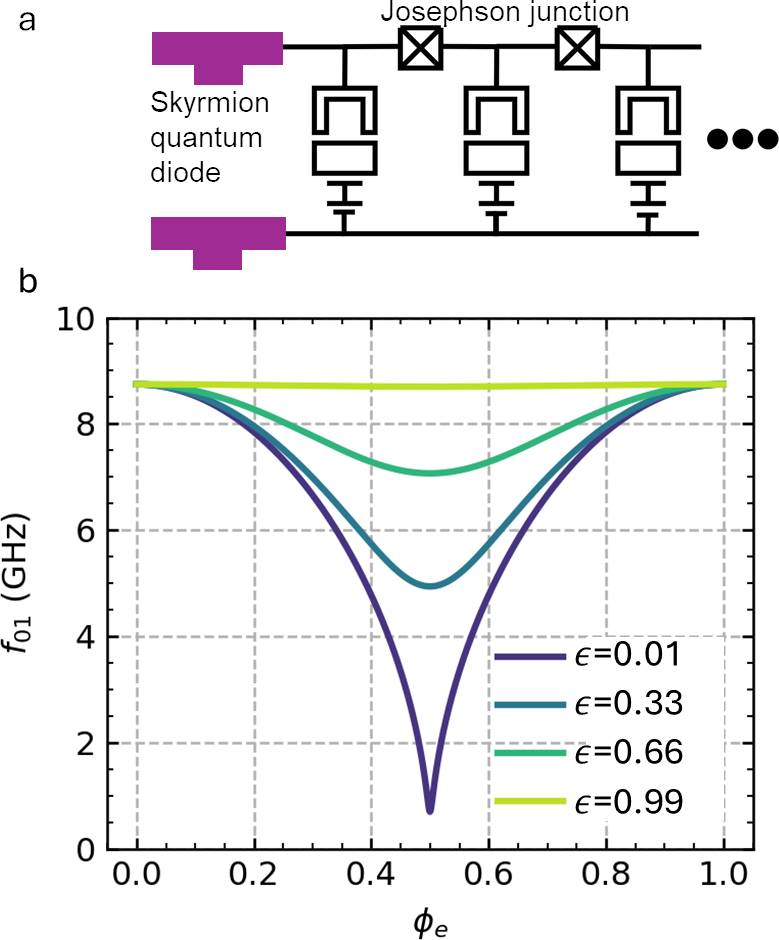}
\caption{\label{fig:tuning}Skyrmion-transmon tuning. (a) Schematic circuit. (b) Transmon qubit resonance frequency plot as a function of $\epsilon$, the imbalance between Josephson energies, and $\phi_e$, the reduced modulation flux.}
\end{figure}

We consider a thin ferromagnetic disk patterned into a T-shaped nanotrack that functions as a skyrmion diode: skyrmions driven from the “input’’ arm propagate unidirectionally toward the “output’’ arm, while reverse motion is suppressed by the combined action of track asymmetry and the skyrmion Hall effect (Fig. \ref{fig:tuning}).  
A flux-tunable transmon chip is mounted directly above the \textit{output} end of the diode.  
The skyrmion’s quantized gyration mode produces a local, time-varying stray-field hotspot concentrated within a few tens of nanometers; at the downstream end this field threads the transmon SQUID loop and modulates its Josephson phase, yielding a magnetic-dipole coupling
\(g_{\mathrm m}\propto(\partial_{\!\varphi}E_J)\,B_\perp(\mathbf r)\) \cite{pan2025tripartite}.  
Because the diode blocks back-propagation, only the forward-moving skyrmion reaches the coupling region, effectively isolating the qubit from upstream noise and reflections.  
The sub-wavelength confinement of the skyrmion reduces participation of dielectric loss channels and provides a compact on-chip mediator for coherent interactions. Further, the setup admits dense tiling of multiple diode–qubit cells on a single chip, paving the way for chiral, scalable quantum buses \cite{jung2021magnetic}.

Then we do flux-dependent resonance frequency derivation. The transmon consists of two Josephson junctions in a SQUID loop threaded by an external flux~\(\phi_e\).  
In the charge basis the Hamiltonian reads
\begin{equation}
  \hat H
  = 4E_C \hat n^{2}
    - E_{J1}\cos\varphi_1
    - E_{J2}\cos\varphi_2,
    \label{eq:3.8}
\end{equation}
with the phase constraint \(\varphi_1-\varphi_2 = 2\pi\phi_e\) and reduced flux
\(\phi_e=\Phi_e/\Phi_0\)\,\cite{koch2007charge, blais2021circuit}.
To eliminate the fast variable, first define symmetric and antisymmetric coordinates
\[
  \varphi = \frac{\varphi_1+\varphi_2}{2},
  \qquad
  \delta  = \frac{\varphi_1-\varphi_2}{2} = \pi\phi_e,
\]
and the total and relative Josephson energies
\[
  E_{J\Sigma}=E_{J1}+E_{J2},
  \qquad
  \epsilon = \frac{E_{J1}-E_{J2}}{E_{J\Sigma}}.
\]
where $\epsilon$ measures the imbalance between the two junctions.
The potential energy becomes
\[
  U(\varphi,\delta)
  = -E_{J\Sigma}\!\left[\cos\delta\,\cos\varphi
        + \epsilon\sin\delta\,\sin\varphi\right].
\]
Minimizing with respect to~\(\delta\) at fixed~\(\varphi\) (valid because \(\delta\) is a fast mode set by the external flux) yields
\begin{equation}
  U(\varphi) = -E_{J,\mathrm{eff}}(\phi_e)\,\cos\varphi,
  \label{eq:3.9}
\end{equation}
where
\begin{equation}
  E_{J,\mathrm{eff}}(\phi_e)
  = E_{J\Sigma}\lvert\cos(\pi\phi_e)\rvert
    \sqrt{1+\epsilon^{2}\tan^{2}(\pi\phi_e)}.
    \label{eq:3.10}
\end{equation}

%\paragraph{Transmon approximation.}
In the regime \(E_{J,\mathrm{eff}}\gg E_C\), expanding the cosine to the fourth
order and treating the quartic term as a perturbation yields the Duffing ladder
\begin{equation}
    \label{eq:3.11}
    \begin{gathered}
    E_m \simeq -E_{J,\mathrm{eff}}
               + \hbar\omega_p\!\Bigl(m+\tfrac12\Bigr)
               - \frac{E_C}{12}\bigl(6m^{2}+6m+3\bigr)\\
    \hbar\omega_p = \sqrt{8E_CE_{J,\mathrm{eff}}}.
    \end{gathered}
\end{equation}

%\paragraph{Fundamental transition.}
The \(m=0\rightarrow1\) spacing is therefore
\[
  E_{01} = \hbar\omega_p - E_C,
\]
so that the flux-dependent resonance frequency becomes
\begin{equation}
  f_{01}(\phi_e)
  = \frac{\sqrt{8E_CE_{J,\mathrm{eff}}(\phi_e)} - E_C}{h}.
  \label{eq:3.12}
\end{equation}
Equations~(\ref{eq:3.10})–(\ref{eq:3.12}) are implemented in the numerical sweep of Fig.~\ref{fig:tuning}, using \(E_{J\Sigma}/h = 50~\text{GHz}\) and
\(E_C/h = 0.20~\text{GHz}\) as is typical of a transmon qubit \cite{koch2007charge}.

%\subsection{Discussion}

The combined analytical results of Equations~(\ref{eq:3.10})–(\ref{eq:3.12}) and the conducted numerical sweep demonstrate, first, that the split-junction transmon offers a seamlessly tunable bridge to the skyrmion’s microwave eigenmode.
Because the qubit resonance frequency $f_{01}(\phi_e)$ varies continuously with the applied SQUID flux, the device can be flux-biased directly onto resonance with the skyrmion’s gyration (or breathing) band or detuned by many linewidths to enter a dispersive regime.
This single-coil control provides an experimentally convenient switch for coherent magnon–photon exchange without the need for additional control circuitry.

The geometry of the T-track diode ensures that only forward-propagating skyrmions reach the coupling region beneath the track’s output arm.
In practice, this means that any broadening, shift, or avoided crossing observed in the qubit spectrum originates unambiguously from magnons arriving via the diode’s preferred direction, while backward-traveling excitations are inherently rejected.
Together, these insights confirm that the hybrid cell not only realizes tunable strong coupling but also embeds directional isolation and straightforward metrology in a compact platform. With no galvanic path, the skyrmion diode also reduces wiring density and lowers the cryostat heat load from attenuated control lines. Its sub-wavelength stray field enables selective addressing of nearby loops, supporting dense layouts with minimal dielectric participation. The skyrmion--transmon cell developed above generalizes naturally to other superconducting qubits \cite{pan2024magnon, lachance2019hybrid, petrovic2025colloquium} because the interaction is, at its core, a flux modulation of a Josephson circuit. For any loop-based qubit with Hamiltonian $H(\Phi)$, the skyrmion’s dynamic stray field produces a small, local flux $\delta\Phi(t)$ threading the loop, and hence an interaction $H_{\text{int}}\!\approx(\partial H/\partial \Phi)\,\delta\Phi(t)$.

At a high level, any superconducting qubit that exhibits a non-vanishing flux susceptibility can interface with the skyrmion diode by positioning a loop or tunable element at the diode’s output end. The diode then acts as a compact, on-chip source of time-dependent magnetic flux that can be tuned into resonance with a target transition or detuned to generate controlled dispersive shifts \cite{blais2021circuit, pioracs2025circuit}.

This mechanism is agnostic to the specific qubit implementation: whether the loop is the qubit’s own (e.g., a flux-type circuit or a capacitively shunted variant) or a nearby tunable element used as a coupler, the Hamiltonian acquires a term proportional to $(\partial H/\partial\Phi)\,\delta\Phi(t)$. Depending on bias, this supports coherent drives, parametric gates, or predominantly longitudinal (dispersive) readout, precisely the modalities engineered in flux-pumped couplers and parametric elements such as Josephson ring modulators (JRMs) and superconducting nonlinear asymmetric inductive elements (SNAILs), now realized magnetically rather than with large on-chip currents \cite{sliwa2015reconfigurable, frattini20173, bergeal2010phase}. A second architectural benefit is directionality. Because the skyrmion track functions as a diode, magnetic excitations propagate one way toward the targeted loop thus suppressing reflections and back-action. In multiplexed layouts, distinct loops can couple to different points along a shared track to achieve spatial fan-out with minimal crosstalk.

From an engineering standpoint, design rules are straightforward: maximize mutual inductance by overlapping the skyrmion stray-field hot spot with the chosen loop; tune geometry/materials so the skyrmion eigenmode lands near the desired microwave transition; and bias for the required interaction character (transverse vs.\ longitudinal) while preserving coherence. Reported skyrmion resonances in multilayers and nanodisks frequently lie in the few- to few-tens-of-GHz range, matching common qubit bands \cite{satywali2021microwave, mochizuki2012spin}.

Looking ahead, the same flux-modulation channel enables reservoir engineering \cite{janovitch2025active} and noise-biasing strategies by coupling selected modes to a directional, lossy port realized by the skyrmion track, complementing conventional cavity or traveling-wave buses. In combination, these ingredients point to compact, nonreciprocal elements that unify control, coupling, and dissipation engineering across a broad class of loop-based superconducting qubits \cite{xia2022nonlinear}.

\section*{Discussion}
We have introduced and analyzed a hybrid architecture that merges a Hall-effect-based skyrmion diode with a flux-tunable superconducting qubit.  
Micromagnetic simulations demonstrate deterministic, one-way skyrmion transport from $\sim20$~nm down to the few-nanometer scale, with sub-nanosecond decision times, while the quantum–circuit model reproduces this nonreciprocal behavior at the qubit level through high forward fidelity and strong isolation. 
The physical model and device concept introduced in Sections: Micromagnetic Simulation of Skyrmion Diode Scaling \& Quantum Simulation with Skyrmion Qubits, which focus on charge-based or position-based skyrmion manipulation in DMI-stabilized systems, and (ii) the qubit encoding schemes discussed in Section: Enhancing Skyrmion Qubit Anharmonicity, which are presented as a broader conceptual extension to alternative material platforms. 
This cross-domain correspondence establishes the physical feasibility of using skyrmions not only as classical information carriers but also as directional, low-loss mediators for quantum signals.
The proposed interface leverages the skyrmion diode’s intrinsic directionality to protect qubits from reverse-propagating noise, while providing flux-based tunability for resonant or dispersive coupling.  

A single nanoscale skyrmion can produce local stray fields (tens to hundreds of mT very close to the surface) that, if brought sufficiently close to a properly designed transmon flux loop or a superconducting pickup, can thread a measurable flux and produce a non-negligible coupling. We can do an estimation by using skyrmion size of $\sim$50 nm and a thickness of 1 nm, which leads to a volume of 8$\times 10^{-24}$ m$^3$. Using a magnetization magnetitude of 10$^6$ A/m, it leads to a magnetic moment of 8$\times 10^{-18}$ A/m$^2$.  Axial dipole field of $\mathbf{m}$ at distance $z$ (on the axis above the skyrmion) is approximately $\mathbf{B_z}\simeq\frac{\mu_0}{4\pi}\frac{2\mathbf{m}}{z^3}$. Considering $z$=50 nm, we have $\mathbf{B_z}$ = 10 mT; $z$=20 nm, $\mathbf{B_z}$ = 200 mT. The skyrmion field is highly localized and the average flux over the whole SQUID loop will be much smaller unless the skyrmion sits directly under a small pickup loop or one junction. Placing a nanoscale pickup loop (area $\sim$ (100 nm)$^2$) right above the skyrmion concentrates the flux coupling and yields far smaller (but more controlled) absolute flux values and it is still easily measurable~\cite{marchiori2022magnetic}. A transmon’s frequency depends on the flux through its SQUID (or flux-bias loop); even small fractions of a flux quantum produce measurable frequency shifts. So a skyrmion that produces a fraction of through a junction or a dedicated pickup coil can shift the transmon frequency and produce dispersive or longitudinal coupling. Several theoretical proposals have shown feasible hybrid coupling schemes between local magnetic textures (including skyrmions) and superconducting circuit~\cite{nothhelfer2022steering,xie2024visualization,menezes2019manipulation}.

Candidate material stacks under consideration include ultrathin magnetic multilayers. For example, zero field isolated skyrmions below 5 nm was observed using spin-polarized scanning tunneling microscopy at 4 K in Rh/Co atomic bilayers on Ir(111)~\cite{meyer2019isolated}. Another example is in our reported Cr$_2$Ge$_2$Te$_6$/Fe$_3$GaTe$_2$ heterostructure~\cite{wu2022van}, the skyrmion size can be as low as 13 nm. Further with ferroelectric tuning~\cite{wu2024room}, the skyrmion size is promised to reach nanoscales. 3 nm skyrmions were also experimentally reported infrustrated triangular-lattice magnet with origin not related to DMI~\cite{kurumaji2019skyrmion}. Compatibility with superconducting circuitry is expected to be acceptable, as skyrmion size decreases at lower temperatures. While stray fields could pose a concern, estimates indicate that an in-plane stray field of $\sim$10 mT from a skyrmion lattice at a distance of 50 nm is unlikely to significantly affect or destroy superconductivity. Also in superconductors like NbSe$_2$, the superconductivity is protected by Ising spin-orbit coupling, preserving its superconductivity even under 9 T in-plane magnetic fields.

Skyrmion diode offers compact footprint and compatibility with room-temperature operation suggest a route to chip-scale, nonreciprocal quantum elements without cryogenic circulators or ferrites.
Future work can target atomistic modeling to validate stability at the smallest sizes, experimental realization of skyrmion-qubit coupling, and integration into larger quantum networks. In parallel, it is promising to carry out experimental studies to demonstrate and quantify the coupling between skyrmions and quantum circuits via transport measurement and magnetic imaging, with the goal of establishing practical skyrmion-qubit interfaces. Meanwhile, scalable integration strategies is still underexplored to incorporate multiple skyrmion-based elements into larger quantum architectures, enabling the development of complex quantum networks and hybrid quantum systems based on skyrmionic devices.
By uniting topologically robust magnetic textures with established superconducting platforms \cite{han2018investigation,wu2019induced}, this approach opens a pathway toward scalable, chiral quantum interconnects and novel hybrid devices operating across classical and quantum regimes.

\section*{Methods}
\subsection*{Micromagnetic simulations}
The simulation of skyrmion motion in the T-shaped track are carried out with MuMax$^3$ software.\cite{vansteenkiste2014design} MuMax$^3$ is a finite element analysis software by solving the Landau-Lifshitz-Gilbert (LLG) equation: 
\begin{equation}
    \frac{\partial\mathbf{m}}{\partial t} = -\gamma_0 \mathbf{m} \times \mathbf{H}_{\mathrm{eff}} + \alpha \left(\mathbf{m} \times \frac{\partial\mathbf{m}}{\partial t}\right) + \bm{\tau}_{\mathrm{SOT}}.
    \label{eq:llg}
\end{equation}

We include the SOT term in the simulation to simulation the skyrmion dynamics under driving current.
We employ the simulation parameters from Fe$_3$GeTe$_2$ hard magnet material and defined size of the material to be $300\times100\times1$ nm$^3$, with the entire region discretized by $1\times1\times1$ nm$^3$. As mentioned in the Section \ref{sec:micromag}, we swept tover the perpendicular anisotropy, and stabilize a skyrmion at $\sim3$nm under the material size at $60\times36\times1$ nm$^3$, discretized by $0.1\times0.1\times1$ nm$^3$. Further discussion regarding SOT term and material parameters are included in Supplementary Note S-I.

\subsection*{Fidelity Numerical simulations}
All simulations were performed in Python using the QuTiP open-quantum-systems library.\cite{lambert2026qutip} Time evolution was computed by numerically solving the Lindblad master equation with the built-in \texttt{mesolve} routine. The system was evolved over a fixed time interval, discretized into 400 time points. For each value of the diode efficiency parameter~$\eta$, the corresponding dissipation rate was assigned and the dynamics were computed independently. Forward and reverse configurations were simulated by initializing the system in the respective basis state and recording the expectation value of the projector onto the target state at each time step. Fidelity maps were constructed by repeating this procedure over a uniformly spaced sweep of~$\eta$. The actual Python implementation were included in Supplementary Note S-III.2.

\section*{Code Availability}
The code for fidelity simulation is available through Supplementary Information. The script for running the micromagnetic simulation is available from corresponding author upon reasonable requestion.

\printbibliography

\section*{Data Availability}

The data are available in UF QESI webpage. The coding is also shown in the Supplementary Information.

\section*{Acknowledgment}
We thank Dr. Chunjing Jia for helpful discussion. This study was funded by UF Gatorade award, Research Opportunity Seed Fund (ROSF) and National Science Foundation ECCS grant No. 2441051. The funders played no role in study design, data collection, analysis and interpretation of data, or the writing of this manuscript.

\section*{Competing Interests}

The authors declare no competing interests.

\section*{Author Contributions}

Y. Wu conceived the ideas and led the project. H. Yang, T. Cao and P. Lu conducted the micromagnetic simulations. G. Bissell performed the anharmonicity analyses and calculations. P. V. Kirk and H. Zhong carried out the fidelity checks. H. Yang and Y. Wu lead the manuscript revision. All authors contributed to writing and revising the manuscript.
\end{document}